\documentstyle[aps]{revtex}
\begin{document}
\draft
\title{Statistical Properties of the Reflectance and Transmittance of an
Amplifying Random Media}
\author{V. Freilikher$^1$, M. Pustilnik$^1$ and I. Yurkevich$^{2,}$\cite{igor}}
\address{$^1$The Jack and Pearl Resnick Institute of Advanced Technology,\\
Department of Physics, Bar-Ilan University, Ramat-Gan 52900, Israel\\
$^2$International Centre for Theoretical Physics, Trieste 34100, Italy}
\date{\today }
\maketitle

\begin{abstract}
Statistical properties of the transmittance ($T$) and reflectance ($R$) of
an amplifying layer with one-dimensional disorder are investigated
analytically. Whereas the transmittance at typical realizations decreases
exponentially with the layer thickness $L$ just as it does in absorbing
media, the average $\left\langle T\right\rangle $ and $\left\langle
R\right\rangle $\ are shown to be infinite even for finite $L$ due to the
contribution of low-probable resonant realizations corresponding to the
non-Gaussian tail of the distribution of $\ln T$. This tail differs
drastically from that in the case of absorption. The physical meaning of
typical and resonant realizations is discussed.
\end{abstract}


Localization of waves in active disordered systems with gain is a subject of
considerable interest given its implications for stimulated emission from
random media [1]. Very recently a number of theoretical [2] - [7] and
experimental [8] - [11] works dealing with this subject have been puplished.
A recent surprising theoretical conclusion is that in one-dimensional random
systems amplification suppresses transmittance just in the same way as
absorption does, so that localization is enhanced due to amplification. In
contrast to this finding in this Letter we show analytically that all
statistical moments $\left\langle T^m\right\rangle $ of the transmittance $T$
of a randomly layered slab with gain are{\it \ infinite} for $m\geq 1$.

Propagation of radiation often exhibits surprising, nonintuitive features
even in homogeneous amplifying systems . Whereas the intensity of radiation
in an infinite active media increases exponentially along the distance of
propagation, the intensity of a plain wave transmitted through a finite
layer as well as the energy of a point source inside the layer exponentially 
{\it decreases} (due to the interference of multiple internal reflection
from boundaries) with the increasing distance . Consider for examle, wave
field $u$ satisfying the one-dimensional stationary Helmholtz equation 
\begin{equation}
\left[ \frac{d^2}{dz^2}+k^2\left( 1+\epsilon (z)\right) \right] u\left(
z\right) =0,  \label{eq1}
\end{equation}
Neglecting nonlinear effects, the amplification of the medium is introduced
by a {\it constant} homogeneous macroscopic parameter - the imaginary part
of the dielectric constant (see discussions in Ref. [3], [4], [7]) $\epsilon
(z)$ $=\varepsilon -i\gamma $ ($\gamma >0$ for amplifying medium) for $0\leq
z\leq L$ ($\epsilon (z)=0$ outside the layer). . The transmittance $T$ and
reflectance $R$ of a layer are expressed through the Green function $%
G_0\left( z,z^{\prime }\right) $ of the wave equation (1) as 
\begin{equation}
T=\left| 2ikG_0\left( 0,L\right) \right| ^2,\;R=\left| 2ikG_0\left(
L,L\right) -1\right| ^2,  \label{eq2}
\end{equation}
where$\;0\leq z\leq z^{\prime }\leq L$, and 
\begin{equation}
G_0\left( z,z^{\prime }\right) =\frac{\left( e^{-i\kappa z}+\rho e^{+i\kappa
z}\right) \left( e^{i\kappa \left( z^{\prime }-L\right) }+\rho e^{-i\kappa
\left( z^{\prime }-L\right) }\right) }{2i\kappa \left( e^{-i\kappa L}-\rho
^2e^{i\kappa L}\right) }  \label{eq3}
\end{equation}
with reflection coefficient $\rho =\left( \kappa -k\right) /\left( \kappa
+k\right) ,$ $\kappa =k\sqrt{1+\varepsilon -i\gamma }\approx k^{\prime
}-i/2l_a,$ $l_a=\left( k\left| \gamma \right| \right) ^{-1}$.

For an arbitrary value of $k^{\prime }=k\sqrt{1+\varepsilon },$ the
denominator in Eq.(\ref{eq3}) is typically a finite (i.e. non-zero) for all $%
L\in \left( 0,\infty \right) $, and when $L/l_a>>1$ the transmittance is
exponentially small {\it independently} on the sign of $\gamma .$
Accordingly, for both absorption ($\gamma <0)$ and amplification $(\gamma
>0) $,

\[
T\propto \exp \left( -L/l_a\right) .
\]
Not only the transmission coefficient, but Green function itself also
exhibits this surprising duality. Indeed, in the limit $z/l_a\gg 1$ and $%
\left( L-z^{\prime }\right) /l_a\gg 1$ (corresponding to a thick layer with
both $z$ and $z^{\prime }$ taken far away from the boundaries), only
contributions from waves reflected from the boundaries (proportional to the
reflection coefficient $\rho $ terms in Eq. (\ref{eq3})) are important in
the Eq.(\ref{eq3}), and Green function $G_0\left( z,z^{\prime }\right) $
(field of a point source) takes the form: 
\begin{equation}
G_0\left( z,z^{\prime }\right) \approx \ \frac{e^{-\left( ik^{^{\prime
}}+1/2l_a\right) \left| z-z^{\prime }\right| \ }}{-2i\kappa \ }=\left( \frac{%
e^{ik^{^{*}}\left| z-z^{\prime }\right| \ }}{2i\kappa ^{*}\ }\right)
^{*} .  \label{eq4}
\end{equation}
Thus from the intensity (either transmitted or inside the layer) there is 
{\it no way of telling what kind of media, amplifying or absorbing, we are
dealing with.}

It was shown recently [5], [7] that an apparent dual symmetry between
amplification and absorption takes place also in random media. In
particular, if the real part of dielectric constant $\varepsilon (z)\ =%
\mathop{\rm Re}
\epsilon (z)$ in Eq. (\ref{eq1}) is a random function of coordinate $z$, and 
$L$ is much larger then the characteristic scattering length $l_s$, the
average specific logarithm of transmittance $\left\langle L^{-1}\ln
T\right\rangle $ is a negative number both for absorbing and amplifying
random media: $\left\langle L^{-1}\ln T\right\rangle =-(1/l_a+1/l_s)$.

We now examine the question: is duality this similarity universal, or it is
a property just of the considered quantity $\left\langle L^{-1}\ln
T\right\rangle $ only? We show in this Letter that generally speaking the
statistics of the intensity of radiation in an amplifying medium differs
drastically from that in an absorbing one. In particular, in systems with
gain all moments $\left\langle T^m\right\rangle $ with $m\geq 1$ are {\it %
infinite} even for finite $L\gg l_s$, while in the case of absorption $%
\left\langle T\right\rangle \propto \exp \left( -L/l_a\right) $, and dies to
zero as $L\longrightarrow \infty .$

To explore the statistical properties of the transmittance in more detail,
we introduce the function 
\begin{equation}
{\cal P}_L^m\left( R\right) =\left\langle T\left( L\right) ^m\delta \left(
R-R\left( L\right) \right) \right\rangle .
\end{equation}
The statistical moments are expressed through this function as

\begin{equation}
\left\langle T^m\right\rangle =\int_0^\infty dR{\cal P}_L^m\left( R\right)
,\;\left\langle R^m\right\rangle =\int_0^\infty dR\,R^m{\cal P}_L^0\left(
R\right) .  \label{eq8}
\end{equation}
To obtain a differential equation for ${\cal P}_L^m\left( R\right) $ we use
the following so-called invariant imbedding equations (see [12] for a useful
introduction) for the complex reflection and transmission coefficients $r$
and $t$ : 
\begin{equation}
\frac{dr}{dL}=2ikr+\frac{ik}2\epsilon (L)\left( 1+r\right) ^2,\;\frac{d\ln t%
}{dL}=\frac{ik}2\epsilon (L)\left( 1+r\right) .  \label{eq5}
\end{equation}
Calculation of $\frac \partial {\partial L}{\cal P}_L^m\left( R\right) {\cal %
\ }$by using Eqs. (5) - (7) and employing the Furutsu - Novikov formula (see
[13]) yields 
\begin{equation}
l_s\frac \partial {\partial L}{\cal P}_L^m\left( R\right) ={\cal L}_m{\cal P}%
_L^m\left( R\right) ,  \label{eq7}
\end{equation}
\[
{\cal L}_m=\frac \partial {\partial R}R\left( \frac \partial {\partial R}%
\left( R-1\right) ^2-2\beta -2m\left( R-1\right) \right) +m^2R+m\left( \beta
-1\right) , 
\]
where $l_s$ is the scattering length in a passive ($\gamma =0$) media and $%
\beta =l_s/l_a$ with $l_a=\left( k\gamma \right) ^{-1}$ being a
characteristic amplification length in the regular ($\varepsilon \left(
z\right) =0$) system. In derivation of Eq. (\ref{eq7}) we have assume that $%
\varepsilon (z)$ is taken to be a Gaussian random process with zero average
and $\left\langle \varepsilon (z)\varepsilon (z^{\prime })\right\rangle
=2\sigma \delta (z-z^{\prime })$ (so that $l_s=2/\sigma k^2$) and that the
rapidly oscillating phase of the reflection coefficient is distributed
uniformly and independently on the reflectance $R=|r|^2$ and the
transmittance $T\ =|t|^{2\text{ }}$(random phase approximation). The initial
and boundary conditions for the Eq.(\ref{eq7}) are: ${\cal P}_{L\rightarrow
0}^m\left( R\right) \rightarrow \delta \left( R-0\right) ,$ and continuity
of ${\cal P}_{L\rightarrow 0}^m\left( R\right) $ at the point $R=1$. The
same equation can be obtained from Fokker-Plank equation of Ref. [\cite
{Beenakker2}] by multiplying it by $T^m$ and integrating over $T$.

Equation (\ref{eq7}) with $\beta $ replaced by $-\beta $ (which corresponds
to absorption), for which $0\leq R\leq 1$, was studied in [13]. It was found
there that the presence of $\beta $ $<0$ leads to spectrum quantization of
the operator ${\cal L}_m$. A similar result occurs in the present case $%
(\beta >0)$ as well. The solution of (\ref{eq7}) may be written in terms of
the eigenstates of ${\cal L}_m$: 
\begin{equation}
{\cal P}_L^m\left( R\right) =\sum e^{-E_n^{(m)}L/l_s}P_n^m\left( R\right) ,
\label{eq9}
\end{equation}
where $P_n^m\left( R\right) $ satisfy the equations 
\begin{equation}
\left( {\cal L}_m+E_n^{(m)}\right) P_n^m\left( R\right) =0.  \label{eq10}
\end{equation}
Keeping only leading terms in the limit $R\rightarrow \infty $ in Eq.(\ref
{eq10}) , one obtains 
\[
\left( \frac \partial {\partial R}R-m\right) ^2RP_n^m\left( R\right)
=0,\;R\rightarrow \infty , 
\]
which gives for the asymptotic form of the $P_n^m\left( R\right) $, 
\begin{equation}
P_n^m\left( R\right) \propto R^{m-2}.  \label{eq11}
\end{equation}
It may be easily seen from (\ref{eq8}), (\ref{eq9}), and (\ref{eq11}) that
all moments $\left\langle T^m\right\rangle $ and $\left\langle
R^m\right\rangle $ {\it diverge} for $m\geq 1$. This results from the fact
that the distribution of transmittance although being almost log-normal near
its maximum [5], [9], changes it drastically in the tail, and decreases like 
$T^{-2}$ for large $T$: 
\begin{equation}
{\cal P}_L\left( T\right) \propto T^{-2},\;T\rightarrow \infty ,
\label{eq12}
\end{equation}
In a similar fashion, the distribution of $R$ behaves as 
\begin{equation}
{\cal P}_L\left( R\right) \propto R^{-2},\;R\rightarrow \infty .
\label{eq13}
\end{equation}
We note that the distribution function for the reflectance $R$ of an
infinite ($L\rightarrow \infty $) amplifying random media has been obtained
in [4], but here we have shown that the average transmittance $\left\langle
T\right\rangle $ and reflectance $\left\langle R\right\rangle $ (as well as
all higher moments of $T$ and $R$) are {\it infinite} for{\it \ }an arbitrary%
{\it \ finite} length $L\gg l_s,l_a$.

The moments $\left\langle T^m\right\rangle $ with $0\leq m<1$ are finite,
and can be calculated explicitly using a suitable modification of approach
of [13]. Let us introduce instead of $R$ in (\ref{eq10}) a new variable $x$
such that $\ R=R_{+}(x)=\left( \tanh x/2\right) ^2$ for $0<R<1,$ and $%
R=R_{-}(x)=\left( \tanh x/2\right) ^{-2}$ for $1<R<\infty $. Introducing new
functions $\Psi _{\pm }^m$ 
\begin{equation}
\Psi _{\pm }^m={\cal P}_L^m\left( R_{\pm }(x)\right) \left| dR_{\pm }\left(
x\right) /dx\right| \exp \left( \Phi _{\pm }^m\left( x\right) /2\right) ,
\label{eq14}
\end{equation}
\[
\Phi _{\pm }^m\left( x\right) =-\ln \left( \sinh x\right) +2m\ln \left( 
\frac 12\left( \cosh x\pm 1\right) \right) \mp \beta \cosh x. 
\]
It may be shown that $\Psi _{\pm }^m$ satisfies a Schr\"odinger equation
with imaginary ''time'' $S=L/l_s$ 
\begin{equation}
-\frac \partial {\partial S}\Psi _{\pm }^m={\cal H}_{\pm }^m\Psi _{\pm
}^m,\;S=L/l_s,  \label{eq15}
\end{equation}
where 
\begin{equation}
{\cal H}_{\pm }^m=-\frac{\partial ^2}{\partial x^2}+V_{\pm }^m\left(
x\right) ,  \label{eq16}
\end{equation}
with the potential 
\begin{equation}
V_{\pm }^m\left( x\right) =\frac 14\left( 1-\sinh ^{-2}x\right) \pm \beta
\left( 1-m\right) \cosh x+\frac 14\beta ^2\cosh ^2x.  \label{eq17}
\end{equation}
The initial and boundary conditions for (\ref{eq15}) follow from that for
Eq.(\ref{eq7}). The spectrum $\left\{ E_n\right\} $ of the operator ${\cal L}%
_m$ is expressed through the spectra $\epsilon _n^{\pm }\left( m\right) $ of
Hamiltonians ${\cal H}_{\pm }^m$ (${\cal H}_{\pm }^m\psi _n^m=\epsilon
_n^{\pm }\left( m\right) \psi _n^m$) as 
\begin{equation}
\left\{ E_n\right\} =\left\{ \epsilon _n^{+}\right\} \cup \left\{ \epsilon
_n^{-}\right\} .  \label{18}
\end{equation}
Note that if $E_n=\epsilon _n^{-}\notin \left\{ \epsilon _n^{+}\right\} $
the corresponding function $P_n^m\left( R\right) $ (see Eq.(\ref{eq7})) is
identically equal zero for $0\leq R\leq 1$. As the random phase
approximation is expected to be valid for small $\beta $ [14], proceeding
along lines of [13] we obtain for $\ln \left( 1/\beta \right) \gg 1$ the
equation 
\begin{equation}
\left( \beta /8\right) ^{2ik_n^{\pm }}=\frac{\Gamma ^2\left( 1+2ik_n^{\pm
}\right) }{\Gamma ^2\left( 1-2ik_n^{\pm }\right) }\frac{\Gamma \left( 1/2\pm
\left( 1-m\right) -ik_n^{\pm }\right) }{\Gamma \left( 1/2\pm \left(
1-m\right) +ik_n^{\pm }\right) },  \label{eq19}
\end{equation}
\[
\epsilon _n^{\pm }=\frac 14+\left( k_n^{\pm }\right) ^2. 
\]

One can see from Eq.(\ref{eq19}) that for integer $m$, $k_n^{+}=k_n^{-}$.
This property is valid for arbitrary $\beta $ also. For $m=0$ it results
from the representation ${\cal H}_{\pm }^0=\chi _{\pm }^{\dagger }\chi _{\pm
}$ with $\chi _{\pm }=\partial /\partial x+\left( 1/2\right) \left( d\Phi
_{\pm }^0/dx\right) $. It is easy to see that $\chi _{+}\chi _{+}^{\dagger
}=\chi _{-}\chi _{-}^{\dagger }$ (both ${\cal H}_{+}^0$ and ${\cal H}_{-}^0$
have the same supersymmetric partner), whence it follows (see [15]) that the
spectra of ${\cal H}_{\pm }^0$ are identical except for the ground states.
For $m=1$ we have ${\cal H}_{+}^0={\cal H}_{-}^0$.

For real $k_n^{\pm }$ (corresponding to $\epsilon _n^{\pm }>1/4$) and $1\leq
n\ll \ln \left( 1/\beta \right) $ Eq.(\ref{eq19}) gives $k_n^{\pm }\approx
\pi n/\ln \left( 1/\beta \right) $, being independent on $m$ in the leading
on $1/\ln \left( 1/\beta \right) $ approximation. For $n=m=0$ we have $%
\epsilon _0^{-}\left( m=0\right) =0$ with exact ground state $\psi _0^0=%
\sqrt{\beta }e^{\beta /2}\exp \left( -\Phi _{-}^0/2\right) ,$corresponding
to stationary distribution ${\cal P}_{L\rightarrow \infty }(R)=2\beta
(R-1)^{-2}\exp \left( -2\beta /\left( R-1\right) \right) $, found in Ref.
[4]). For small, but finite $m$ Eq.(\ref{eq19}) yields 
\begin{equation}
\epsilon _0^{-}\left( m\right) =m\left( 1+\beta \right) +m^2\left( 1-2\beta
\ln 1/\beta \right) +O(m^3)  \label{eq20}
\end{equation}
The term linear in $m$ in this expansion can be easily reproduced for
arbitrary $\beta $ by perturbation theory: $\epsilon _0^{-}\left( m\right)
=\left\langle \psi _0^0\left| V_{-}^m-V_{-}^0\right| \psi _0^0\right\rangle
=m\left( 1+\beta \right) $.

If $L/l_s\gg 1,$ only contribution from the lowest level $E=\epsilon _0^{-}$
in the sum in (9) is important, and from (\ref{eq20}) and (\ref{eq9}) we
obtain for $0\leq m<1:$

\begin{equation}
\left\langle T^m\right\rangle \propto \exp \left( -\epsilon _0^{-}\left(
m\right) L/l_s\right) .\;\   \label{eq21}
\end{equation}
Taking into account that $\left\langle \ln ^nT\right\rangle =\left( \partial
^n\left\langle T^m\right\rangle /\partial m^n\right) _{m\rightarrow 0}$, and
using Eqs. (\ref{eq20}) and (\ref{eq21}) we get:

\begin{equation}
\left\langle \ln T\right\rangle =-(1-\beta )L/l_s,\;%
\mathop{\rm var}
(\ln T)=2\left( 1-2\beta \ln 1/\beta \right) ,  \label{eq22}
\end{equation}
in agreement with the results of [7] (see also [16]).

Formulas (\ref{eq22}) correspond to the exponential decay of the
transmittance of a random amplifying layer that at first glance seems to
contradict to the infinite value of $\left\langle T\right\rangle $ obtained
above. In fact these two results are actually in agreement. Since (in
accordance with Eq.(\ref{eq22})) fluctuation of $\xi =L^{-1}\ln T$ decays to
zero when $L\rightarrow \infty ,$ $\xi $ is a self-averaged random quantity
whose value at each random realization tends (with the probability one) to
the non-random limit $\xi _{L\rightarrow \infty }=-(1-\beta )/l_s$ which
corresponds to the maximum of the distribution function $P(\xi )$. Therefore
the exponentially decreasing function $e^{\left\langle \ln T\right\rangle
}=e^{-(1-\beta )L/l_s}$ represents the behavior of the random quantity $T$
at the most probable (typical) realizations. At the same time the main
contribution to the average transmittance $\left\langle T\right\rangle $
comes from rare, low-probable (so-called representative) realizations
(corresponding to the far tail of the distribution function $P(\xi )$),
where the transmittance takes (due to resonances) an infinite value.

The existence of two types of realizations (typical and resonant) can be
demonstrated by the simplest example of a random ensemble of amplifying
systems: Fabry-P\'erot resonators (dielectric layers) with homogeneous at
each realization, but randomly fluctuating from sample to sample width $L$
and real part of dielectric constant $\varepsilon (z).$ As was shown above,
the transmittance of a dielectric layer is exponentially small independent
of the sign of $\gamma $, i.e. both for absorption and amplification.
However, in the case of amplification ($\gamma <0$) there exist {\it a} {\it %
discrete }set of resonant pairs $\{k_n^{\prime },\ \ L_n\}$ for which the
denominator of $\ $Eq. (\ref{eq3}) equals zero, and $G_0\left( z,z^{\prime
}\right) $ as well as the transmittance $T$ and reflectance $R$ become
infinite. The occurrence of such resonant realizations may give rise to the
divergences of $\left\langle T\right\rangle $ and $\left\langle
R\right\rangle $. Mathematically, if we first average $T$ (and $R$) over the
rapid phase oscillations related to the wave length $2\pi /k^{\prime }\ll
l_a $ (this is equivalent to the random-phase approximation used in the
derivation of Eq.(\ref{eq7})), and then average the resulted ''smoothed ''
functions over random $\varepsilon $, we obtain (due to the infinite
contributions of the above-mentioned resonant realizations) infinite values
for $\left\langle T\right\rangle $ and $\left\langle R\right\rangle $.$\ $

To conclude, statistical moments of the transmittance $\left\langle
T^m\right\rangle $ and reflectance $\left\langle R^m\right\rangle $ for a
disordered amplifying layer of the thickness $L$ are calculated
analytically. When $m<1$ moments $\left\langle T^m\right\rangle $ as well as 
$\left\langle \ln T\right\rangle $ and $%
\mathop{\rm var}
\ln T$ exhibit the same $L$-dependance as in absorbing ($%
\mathop{\rm Im}
\epsilon \longrightarrow -%
\mathop{\rm Im}
\epsilon $) media. This behavior is inherent for {\it typical }realizations%
{\it \ }corresponding to the Gaussian-like part of the $\ln T$ distribution
function near its maximum. However, the duality between absorbing and
amplifying media breaks down for $m\geq 1$: all moments $\left\langle
T^m\right\rangle $ and $\left\langle R^m\right\rangle $ with $m\geq 1$
diverge (due to the contribution of low-probable resonant realizations
related to the non-Gaussian tail of the distribution) for finite $L$ that is
larger than the characteristic scattering length $l_s.$

We are grateful to I. Freund for stimulating discussions and useful
criticism.


\begin{references}
\bibitem[*]{igor}  on leave from the Institute of Low Temperature Physics
and Engineering, Kharkov 310164, Ukraine

\bibitem{Let}  M. Kempe, G. A. Berger, and A. Z. Genack, in ''{\it Handbook
of Optical Properties''}, vol.II, Ed. by R.E. Hummel and P. Willmann, CRC
Press, Boca Raton, 1996.

\bibitem{Z}  A. Yu. Zyuzin,\ Europhys. Lett. {\bf 26}, 517 (1994), JETP
Lett. {\bf 61}, 990 (1995).

\bibitem{Z1}  A. Yu. Zyuzin,\ Phys. Rev. E {\bf 51}, 5274 (1995).

\bibitem{K}  P. Pradhan and N. Kumar, Phys. Rev. B {\bf 50}, 9644 (1994).

\bibitem{Zhang}  Z. Q. Zhang, Phys. Rev. B {\bf 52}, 7960 (1995).

\bibitem{Beenakker1}  C. W. J. Beenakker, J. C. J. Paasschens, and P. W.
Brouwer, preprint cond-mat/9601024.

\bibitem{Beenakker2}  J. C. J. Paasschens, T. Sh. Misirpashaev, and C. W. J.
Beenakker, preprint cond-mat/9602048.

\bibitem{E1}  A. Z. Genack and J. M. Drake, Nature {\bf 368}, 400 (1994).

\bibitem{E2}  A. N. Lawandy, R. M. Balachandran, A. S. L. Gomes, and E.
Sauvaln, Nature {\bf 386}, 436 (1994).

\bibitem{Sha}  W. L. Sha, C. -H. Liu, and R. R. Alfano, Opt. Lett. {\bf 19},
1922 (1994).

\bibitem{E3}  D. S. Wiersma, M. P. van Albada, and A. Lagendijk, Phys. Rev.
Lett. {\bf 75}, 1739 (1995).

\bibitem{RD}  R. Rammal and B. Doucot, J. Physique {\bf 48}, 509 (1987).

\bibitem{We}  V. Freilikher, M. Pustilnik, and I. Yurkevich, Phys. Rev.
Lett. {\bf 73}, 810 (1994); Phys. Rev. B {\bf 50}, 6017 (1994).

\bibitem{Y}  A.K. Gupta and A.M. Yayannavar, Phys. Rev. B{\bf \ 52}, 4156
(1995).

\bibitem{SUSY}  F. Cooper, A. Khare, and U. Sukhatme, Phys. Rep. {\bf 251},
267 (1995).

\bibitem{lan}  $\left\langle \ln T\right\rangle $ can be derived directly
from Eq. (\ref{eq5}). After a simple algebra one gets $l_sd\ln T/dL=l_sd\ln
\left| R-1\right| /dL-\beta \left| 1+r\right| ^2/\left( R-1\right) $.
Neglecting oscillating terms and averaging with stationary distribution of $%
R $, one obtains $\left( l_s/L\right) \left\langle \ln T\right\rangle
=\left\langle \beta \left( R+1\right) /\left( R-1\right) \right\rangle
=-\left( 1+\beta \right) $.
\end{references}
\end{document}